\documentclass[prb,showpacs,twocolumn,citeautoscript,superscriptaddress]{revtex4}

\usepackage{amsmath,graphicx,latexsym,bm}

\begin{document}

\title{Berry-phase theory of polar discontinuities at oxide-oxide interfaces}

\author{Massimiliano Stengel}
\affiliation{CECAM - Centre Europ\'een de Calcul Atomique et Mol\'eculaire,
Ecole Polytechnique F\'ed\'erale de Lausanne, 1015 Lausanne, Switzerland}
\author{David Vanderbilt}
\affiliation{Department of Physics and Astronomy, Rutgers University, Piscataway,
             New Jersey 08854-8019, USA}
\date{\today}

\begin{abstract}
In the framework of the modern theory of polarization, we rigorously
establish the microscopic nature of the electric displacement field
$\mathbf{D}$.
In particular, we show that the longitudinal component of ${\bf D}$ is
preserved at a coherent and insulating interface.
To motivate and elucidate our derivation, we use the example of
LAO/STO interfaces and superlattices, where the validity of
the above conservation law is not immediately obvious.
Our results generalize the ``locality principle'' of constrained-${\bf D}$
density functional theory to the first-principles modeling of charge-mismatched systems.
\end{abstract}

\marginparwidth 2.7in
\marginparsep 0.5in
\def\msm#1{\marginpar{\small MS: #1}}
\def\dvm#1{\marginpar{\small DV: #1}}
\def\scr{\scriptsize}

\pacs{71.15.-m, 77.22.Ej, 77.55.+f, 77.84.Dy, 73.61.Ng}

\maketitle


In classical electrostatics, the normal component of the electric
displacement field ${\bf D}$ is preserved,
\begin{equation}
({\bf D}_1 - {\bf D}_2)\cdot {\bf \hat{n}} = 0 \, , 
\label{eqdd}
\end{equation}
at an insulating (charge-carrier-free) planar interface between
two homogeneous insulators.
Here ${\bf D}_1$ and ${\bf D}_2$ are the values of the macroscopic electric
displacement in material 1 and 2 and ${\bf \hat{n}}$ is the plane
orientation.
While this is in principle a macroscopic law, recent first-principles
calculations (e.g., on multicomponent perovskite
superlattices\cite{xifan_sl}) have shown that it is applicable even
at the \emph{microscopic} level.
Given this success, one would be tempted to adopt Eq.~(\ref{eqdd}) in
full generality for the description of insulating interfaces at the nanoscale,
where the conservation of $D = {\bf D}\cdot {\bf \hat{n}}$
facilitates the interpretation and modeling of many phenomena.

There are, however, a number of cases which have emerged recently
where the applicability of Eq.~(\ref{eqdd}) is not immediately obvious.
The prototypical example is that of a charge-mismatched
interface between two crystalline insulators.
Polar interfaces have been the object of intense research in the past
few years, motivated by the recent observation of
two-dimensional metallicity at interfaces between LaAlO$_3$ (LAO)
and SrTiO$_3$ (STO)~\cite{Nakagawa:2006,Ohtomo:2004}.
Interestingly, first-principles calculations have recently demonstrated
that the LAO/STO interface can remain \emph{insulating} under certain
conditions, and in such a regime the oxide lattice undergoes rather unusual
relaxation patterns.
For example, in the case of a thin film of LAO deposited on the
(001) surface of STO, strong ``ferroelectric-like'' polar
distortions were reported in the LAO overlayer while the substrate
remains essentially unperturbed~\cite{pentcheva}.
Conversely, in periodic LAO/STO superlattices it was shown that the
LAO and STO components spontaneously polarize in \emph{opposite}
directions, the largest distortions occurring now in STO~\cite{bristowe}.
The source of this polarization discontinuity $\Delta P$ is understood:
it is induced by electric fields arising from extra interface charges
$\pm e/2$ (for IV-III and II-III interfaces respectively)
resulting from the ``polar discontinuity'' between the II-IV
(STO) and III-III (LAO) perovskites.
Similar issues arise for the case of II-IV/I-V (e.g.,
STO/KNbO$_3$) superlattices~\cite{eamonn_kn}.
In these examples one is seemingly forced to conclude that
Eq.~(\ref{eqdd}) is violated, since the $\Delta P$ gives rise
to a corresponding discontinuity in $D$ that is inconsistent
with Eq.~(\ref{eqdd}).

One way to resolve this issue is to accept a definition of
``free charge'' as including {\it everything}
other than bound charge, as suggested by some authors (e.g.,
Ref.~~\onlinecite{griffiths}, Sec.~4.3.1).
Then Eq.~(\ref{eqdd}) is fixed by adding a ``free charge''
term on the right to
represent the polar-discontinuity charge, even though this
is anything but ``free.''
Another possibility, proposed by
Murray and Vanderbilt~\cite{eamonn_kn}, is to introduce
a \emph{compositional} charge density
$\rho_{\rm comp}$, which is distinct both from bound
charge ($\rho_{\rm bound}=-\nabla\cdot \mathbf{P}$) and
from truly free charge (associated with charge carriers), and
write
\begin{equation}
\rho_{\rm tot} = \rho_{\rm free} + \rho_{\rm bound} + \rho_{\rm comp} .
\label{eqrho}
\end{equation}
With $\nabla\cdot\mathbf{D}=4\pi(\rho_{\rm free}+\rho_{\rm comp})$,
Eq.~(\ref{eqdd}) becomes, for an insulating interface,
\begin{equation}
({\bf D}_1 - {\bf D}_2)\cdot {\bf \hat{n}} = 4 \pi \sigma_{\rm comp}.
\label{eqrhocomp}
\end{equation}

While these approaches already provide a satisfactory account
of the phenomena described in Refs.~~\onlinecite{pentcheva,bristowe,eamonn_kn},
there are good reasons to seek an alternative description.
First, both of the above approaches are somewhat awkward, either
introducing a third kind of charge, or describing as ``free'' a form
of charge that is fixed.
Second, it is necessary to assess the range of applicability
of Eq.~(\ref{eqdd}), i.e. to identify a general criterion for
deciding in what cases it should be replaced by Eq.~(\ref{eqrhocomp}).
Third, one would like to establish a true microscopic definition of 
all quantities appearing in Eq.~(\ref{eqrho}), particularly
$\rho_{\rm comp}({\bf r})$ and $\rho_{\rm bound}({\bf r})$.
Given the ever-increasing interest in perovskite thin films and
superlattices, such an analysis is particularly urgent in order
to provide timely support for the experimental work with
appropriate modeling tools.

Here we show that the incompatibility between Eq.~(\ref{eqdd})
and the ``polar discontinuity'' can instead be elegantly
resolved in the framework of the Berry-phase modern theory of polarization.
In particular, we show that Eq.~(\ref{eqdd}) is exact without the need
for extensions, once $D$ is expressed in terms of the formal
(rather than the effective) macroscopic polarization, and is
a direct consequence of the interface theorem~\cite{Vanderbilt/King-Smith:1994}.
This result puts the ``locality principle'' of constrained-$\mathbf{D}$
density functional theory~\cite{xifan_sl} on a firm theoretical basis,
and generalizes its scope to the case of charge-mismatched
superlattices and heterostructures.
In the following, we will first introduce some basic properties
of the formal polarization in periodic insulators.
Based on these properties, we will then establish
the link between Eq.~(\ref{eqdd}) and the interface theorem.
Finally, we will demonstrate these ideas in practice by performing
calculations of a model LAO/STO superlattice.

\begin{figure}
\includegraphics[width=3.2in]{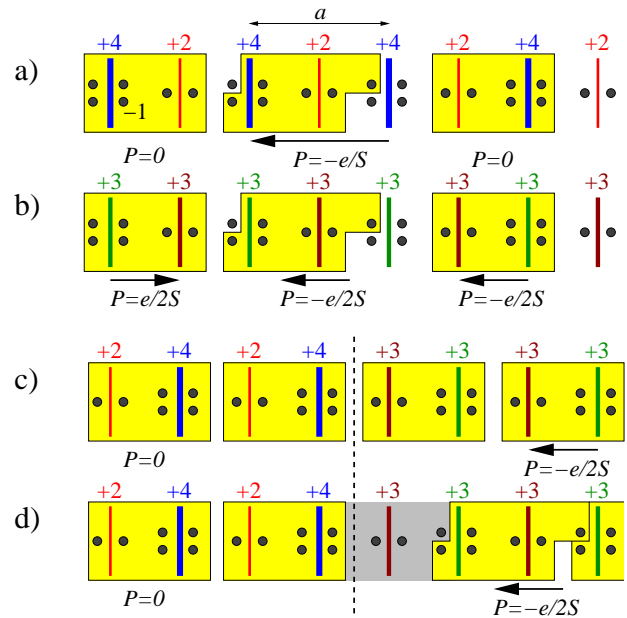}
\caption{ (Color online)
Schematic representation of formal polarization in the bulk
(a-b) and at interfaces (c-d).  Blue and red vertical
lines represent BO$_2$ and AO layers in a II-IV perovskite (a);
green and brown are the same but in a III-III perovskite (b).
Black circles represent electronic Wannier centers
(not all are explicitly shown).  Different choices of basis
(grouping) lead to values of $P$ differing by $\Delta P = e/S$.
(c) Choice of basis such that no charges are left in the interface
region.  (d) Choice of bias leaving net-neutral interface region
(grey).
\label{figquantum} }
\end{figure}

In an independent-electron (or mean-field) picture, the total charge
density of an insulator can be expressed as a sum of contributions from
ion cores and Wannier functions.
Let ${\bf r}_\alpha$ and ${\bf r}_i$ represent the ion core and
Wannier center locations, respectively, for a choice of ``basis''
that tiles the crystal under primitive translations ${\bf a}_{1}$,
${\bf a}_{2}$ and ${\bf a}_{3}$.
The dipole moment of this discrete set of charged objects divided by the
cell volume defines the macroscopic polarization 
${\bf P}$~\cite{King-Smith/Vanderbilt:1994,Resta_rmp}.
We shall be concerned with a
single component of the polarization,
i.e., its projection in the direction $\bf\hat{n}$ parallel to
${\bf a}_2\times{\bf a}_3$; let this be direction $x$.
Then we have an essentially one-dimensional problem with $P=P_x$
given by
\begin{equation}
P = \frac{1}{\Omega} \Big(
\sum Q_\alpha x_\alpha -
e \sum x_i \Big) \, ,
\label{eqformalp}
\end{equation}
where $Q_\alpha$ and $-e$ are the charges of a given ionic core and
Wannier function, respectively, and $\Omega$ is the cell volume.
While this is rigorous, it has a degree of arbitrariness
in that one may choose to include in the crystal basis any of the
infinitely repeated images of each Wannier function or ion core.
As shown schematically in Fig.~\ref{figquantum}(a), this implies
that $P$ is a multi-valued function; it is only defined modulo a
``quantum of polarization'' $\Delta P= e/S$, where
$S=|{\bf a}_2 \times {\bf a}_3|$ is the cell surface area.

The polarization in Eq.~(\ref{eqformalp}) is the \emph{formal}
polarization corresponding to the raw result of a
Berry-phase calculation (in contrast to
the \emph{effective} polarization, defined
relative to a centrosymmetric reference)~\cite{ferro:2007}.
It is essential to understand that the formal polarization does
\emph{not} necessarily vanish in a centrosymmetric material
(i.e., $\bf P$=0 is not contained in the lattice of allowed
values).
Practical realizations of this situation are III-III perovskites like
LAO in their cubic five-atom reference structure.
With $x$ along the (100) direction, the lattice of allowed
values is $P=\pm e/2S,\;\pm3e/2S,\;...$, as shown schematically
in Fig.~\ref{figquantum}(b).
This occurs essentially because the individual LaO and AlO$_2$
layers have formal charges of $\pm e$.
In contrast, II-IV perovskites like STO, which have formally
neutral AO and BO$_2$ layers, have
$P=0,\,\pm e/S,\;...$, as illustrated in Fig.~\ref{figquantum}(a).
We now show that this fundamental difference in the respective
$P$-lattices of LAO and STO is the key to reconciling Eq.~(\ref{eqdd})
and the ``misbehavior'' of charge-mismatched interfaces.

Consider a coherent insulating~\cite{explan-insulating}
interface between LAO and STO
with perfect $(1\times 1)$ periodicity, i.e. ${\bf a}_2$ and ${\bf a}_3$
are common to both crystals and lie on the interface plane.
Fig.~\ref{figquantum}(c-d) shows an example with a TiO$_2$:LaO interface
termination, but similar considerations apply for other terminations.
Then it is always possible to group ions and Wannier functions
into a basis for STO and one for LAO (yellow areas in the figure)
such that the left-over
interface region (grey shaded area) is overall
charge neutral; two examples are provided in the figure.
We use these basis conventions to define the bulk polarization
in STO ($P_1$) and LAO ($P_2$) via Eq.~(\ref{eqformalp}).
Because of the neutrality of the interface region, it follows
that the macroscopic density of \emph{bound} charge at the interface is
\begin{equation}
\sigma_{\rm bound} = P_1 - P_2 .
\label{eqintth}
\end{equation}
This is essentially the ``interface theorem'' of
Ref.~~\onlinecite{Vanderbilt/King-Smith:1994}, a central result of the
modern theory of polarization.

Eq.~(\ref{eqintth}) implies that the bound charge at the interface
is determined by the discontinuity of \emph{formal} bulk
polarizations of the two
participating materials, removing the need to complement
the theory with $\rho_{\rm comp}$~\cite{explan-delta}.
At the TiO$_2$:LaO interface sketched in
Fig.~\ref{figquantum}(c-d), for example, Eq.~(\ref{eqintth}) would
yield a $\sigma_{\rm bound}$ of exactly half an electron per unit cell
if both bulks were centrosymmetric,
consistent with the heuristic arguments of the ``polar 
catastrophe''~\cite{Ohtomo:2004} model.
However, macroscopic electric fields $\mathcal{E}_1$ and $\mathcal{E}_2$
will typically be present in materials 1 and 2, since
Gauss's theorem implies that
\begin{equation}
4\pi \sigma_{\rm bound} =
-\mathcal{E}_1 + \mathcal{E}_2 \, ,
\label{eqfield}
\end{equation}
and the self-consistent solution is the one in which the polarizations
$P_1$ and $P_2$ are the equilibrium values in the corresponding
fields $\mathcal{E}_1$ and $\mathcal{E}_2$.
These polar distortions, in turn, screen the discontinuity in $P$.
Introducing the electric displacement~\cite{fixedd}
\begin{equation}
{\bf D} = \bm{\mathcal{E}} + 4\pi {\bf P} \, ,
\label{eqd}
\end{equation}
it follows from Eq.~(\ref{eqfield}) that $D_1 - D_2 = 0$, which is
exactly Eq.~(\ref{eqdd}).
[Note that the multivalued character of $P$ propagates to the
$D$ through Eq.~(\ref{eqd}); $D$ is lattice-valued
with a ``quantum'' $\Delta D=e/4\pi S$.]
Thus, the phenomena discussed in the introduction emerge as
a \emph{consequence} to Eq.~(\ref{eqdd}), rather than as an exception to it.

To substantiate this interpretation, where all charge not associated
with free carriers is counted as bound charge,
$\rho_{\rm bound} (\mathbf{r}) = \rho_{\rm tot} (\mathbf{r})$,
it is useful to look back at the theory developed in
Ref.~~\onlinecite{Giustino/Pasquarello:2005}.
The authors defined a ``local'' polarization through~\cite{explan-constant}
\begin{equation}
\frac{d \bar P(x)}{dx} = -\bar{\rho}_{\rm bound}(x) ,
\label{eqdp}
\end{equation}
where the bar indicates planar averaging over the $yz$ planes
and ``macroscopic'' averaging~\cite{macroscopic} along the
heterostructure stacking direction $x$.
This leads to a microscopic definition of the displacement field,
$\bar D(x)=\bar{\mathcal{E}}(x) + 4\pi \bar P(x)$, where the local
electric field is given by the \emph{microscopic} Maxwell equations,
\begin{equation}
\frac{d\bar{\mathcal{E}}(x)}{dx} = \frac{1}{4\pi} \bar{\rho}_{\rm tot}(x) .
\label{eqde}
\end{equation}
Combining Eqs.~(\ref{eqdp}) and (\ref{eqde}) we obtain that $\bar D(x)$ must
be constant throughout the superlattice, consistent with
Eq.~(\ref{eqdd}). It is also straightforward to show that
$\bar P(x)$ converges to $P_1$ or $P_2$ sufficiently
far from the interface.

These arguments, therefore, generalize both the ``locality principle'' of
constrained-$D$ theory~\cite{xifan_sl} and the theory of
local dielectric response~\cite{Giustino/Pasquarello:2005}
to the case of charge-mismatched systems.
However, this result comes at a price: One must accept that the
formal polarization in a centrosymmetric perovskite
like LAO does not vanish (and might even be non-zero in centrosymmetric
STO, depending on the branch choice).
While this may be counterintuitive, it is an established
aspect of the modern theory of polarization~\cite{ferro:2007},
and has been already crucial for answering important questions in the physics
of complex oxides such as BiFeO$_3$~\cite{Neaton_et_al:2005}.
Here we show that the ``half-quantum'' nature of a III-III (or I-V)
perovskite is not merely a technical annoyance; rather, it
acquires a well-defined physical meaning through Eq.~(\ref{eqintth}).

\begin{figure}
\includegraphics[width=3.2in]{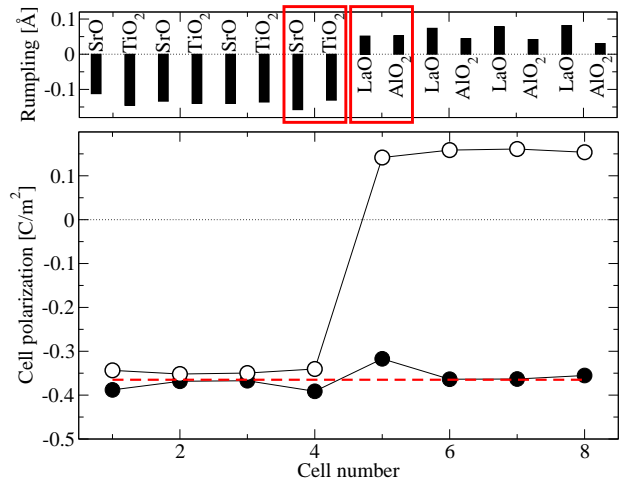}
\caption{(Color online) Top panel: cation-oxygen rumplings, $\delta_{\rm AO}$ and $\delta_{\rm BO2}$,
in a given layer of the (LAO)$_4$/(STO)$_4$ superlattice.
The red boxes indicate the
grouping of the layers adopted for defining $P_j$ and $\tilde P_j$.
Bottom panel: cell polarizations $\tilde P_j$ (open symbols),
$P_j$ (filled symbols) and macroscopic value $P_{\rm Berry}$
(dashed horizontal line).}
\label{fig1}
\end{figure}

To elucidate in practice the continuity of the formal polarization
at a polar interface, we now consider explicitly a periodic superlattice
composed of four layers each of STO and LAO in a
tetragonal $1 \times 1$ supercell with the in-plane lattice
parameter set to the theoretical equilibrium value of
$a_0$ = 7.274 a.u. 
for cubic STO.
Our calculations are performed within the local-density approximation
of density-functional theory and the projector-augmented-wave
method \cite{Bloechl:1994}, with a plane-wave cutoff of 60 Ry.
The Brillouin zone is sampled with a $(6\times 6 \times 1)$
Monkhorst-Pack grid.
We relax the ionic forces and the out-of-plane stress
to a tolerance of $10^{-5}$ Ha/bohr and $10^{-7}$ Ha/bohr$^2$,
respectively.
After relaxation the system remains insulating with a large gap of
$\sim$1.5 eV, in qualitative agreement with Ref.~~\onlinecite{bristowe}.
We then inspect the local polarization profile using two
contrasting approaches as follows.

First, we define a local \emph{formal} polarization $P_j$ for each
of the eight cells $j$ by inserting the coordinates of the ionic
positions and Wannier centers of the cell in question into
Eq.~(\ref{eqformalp})~\cite{explan-volume}.
The definition of a ``cell'' is fixed by the previous
choice of basis for each bulk material.
The calculated $P_j$ values are shown as the filled circles in
Fig.~\ref{fig1}, where it is assumed that the chosen basis was
the one sketched in the top panel of Fig.~\ref{fig1}, also
corresponding to Fig.~\ref{figquantum}(c).
The total Berry-phase polarization of the supercell, 
$P_{\rm Berry}= 1/8 \sum_j P_j$, is 
then simply the average of the local $P_j$ values. 
(Our definition of $P_j$ is closely related to the prescription of
Ref.~~\onlinecite{xifan_lp}, except that a further subdivision
into contributions from AO and BO$_2$ layers was introduced there.)

Second, by using the same grouping convention, we define the local
\emph{effective} polarization $\tilde P_j$ of cell $j$ as
\begin{equation}
\tilde P_j = \frac{e}{\Omega} \sum_{\alpha \in j} Z^*_\alpha \Delta x_\alpha \, ,
\end{equation}
where $Z^*_\alpha$ and $\Delta x_\alpha$ are the Born effective charge
and atomic displacement of atom $\alpha$ respectively.
(Displacements are defined relative to a stack of ideal centrosymmetric
cells.)
The resulting $\tilde P_j$ values are plotted as the open symbols in
Fig~\ref{fig1}.

Strikingly, the Wannier-based formal polarization $P_j$ is rather
uniform, with little departure from the average value $P_{\rm Berry}
=-0.365$ C/m$^2$ shown as the dashed line in Fig.~\ref{fig1}.
This confirms the validity of our arguments:
At electrostatic equilibrium the system tries to minimize the
macroscopic electric fields present on either side of the
interface by making the discontinuity in the ``formal'' $P$ as
small as possible.
Interestingly, in our heterostructure, the LAO and STO components
acquire oppositely oriented structural distortions (upper panel of
Fig.~\ref{fig1}) in order to achieve this goal.
This is reflected by the abrupt discontinuity in the ``effective''
local polarization pattern $\tilde P_j$.
Note that in the STO region $\tilde P_j$ and $P_j$ almost coincide,
since our branch choice for the formal polarizations implies $P=\tilde P =0$
for bulk STO in its centrosymmetric ground state.
The difference between $\tilde P_j$ and $P_j$ in the LAO
region of the plot (Fig.~\ref{fig1}) corresponds to half a quantum of
polarization, as expected.

To corroborate our conclusions, we now show that
the local properties deep in the LAO and STO regions are determined
by the macroscopic $D$, and do not directly depend on the
specific details of the interface.
To that end, we calculate (within the same computational
parameters and symmetry constraints) the equilibrium structure
of bulk LAO and bulk STO with $D$ constrained~\cite{fixedd}
to take the value $D$=$-$0.365\,C/m$^2$
extracted from the heterostructure in electrostatic equilibrium.
The average bucklings
$\delta_{\rm LaO}$=0.078\,\AA\ and $\delta_{\rm AlO2}$=0.044\,\AA\ extracted
from cells 6-7 are in excellent agreement with bulk LAO values of
0.078\,\AA\ and 0.043\,\AA\ 
respectively, while $\delta_{\rm SrO}$=$-$0.140\,\AA\ and
$\delta_{\rm TiO2}$=$-$0.140\,\AA\ (coincidentally the same for both layers)
extracted from cells 2-3 
similarly match the bulk STO values of $-$0.137\,\AA\ 
and $-$0.140\,\AA{}.
The largest discrepancy is $\sim$3\,m\AA, confirming the soundness of
our arguments.
A similar reasoning can be used to interpret the results of
Refs.~~\onlinecite{pentcheva} and ~\onlinecite{Lee/Demkov:2008}:
when $D$=0 is enforced by symmetry, an electric field~\cite{Lee/Demkov:2008} of
$\mathcal{E}$=0.24\,V/\AA{} and significant ferroelectric-like
rumplings~\cite{pentcheva} ($\delta_{\rm LaO}$=0.26\,\AA, 
$\delta_{\rm AlO2}$=0.15\,\AA) were reported in the LAO layer.
To check whether these results are consistent with our arguments,
we repeated our \emph{bulk} LAO calculations at $D$=0 and found
$\mathcal{E}$=0.238\,V/\AA, $\delta_{\rm LaO}$=0.241\,\AA\ and
$\delta_{\rm AlO2}$=0.143\,\AA, in excellent agreement with the
literature values.

In conclusion, we have demonstrated that the conservation of
the longitudinal component of $D$ expressed in Eq.~(\ref{eqdd}) is
a \emph{microscopic} law that explains and predicts the
behavior of insulating interfaces and superlattices in a variety
of electrical boundary conditions.
This fundamental principle, together with state-of-the art
finite field approaches~\cite{fixedd,xifan_sl}, is a powerful
theoretical tool to complement and guide experiments
in the emerging field of interface nanoengineering.
Our ideas are very general, and can readily be applied to a wide
range of physical systems such as, e.g., nitride superlattices
where the issue of interface polarity poses important technological
challenges~\cite{speck_mrs}.

M.S. thanks N. A. Spaldin for stimulating discussions. 
This work was supported by ONR grant N00014-05-1-0054 (D.V.).
Calculations were performed at NCSA.

\bibliography{Max}

\begin{thebibliography}{23}
\bibitem[{exp({\natexlab{a}})}]{explan-insulating}
\bibinfo{note}{This statement requires some care when a macroscopic electric
  field is present in one or both semicrystals. We assume that the two
  participating materials are thick enough for the local interface distortions
  to heal completely, but thin enough to avoid dielectric breakdown.}
\bibitem[{exp({\natexlab{b}})}]{explan-delta}
\bibinfo{note}{In some cases, as for $\delta$-doping layers with a random
  planar distribution of dopants, it might still be more natural to work with
  $\rho_{\rm comp}$; it is important to stress the requirement of in-plane
  periodicity in the derivation of Eq.~(\ref{eqintth}).}
\bibitem[{exp({\natexlab{c}})}]{explan-constant}
\bibinfo{note}{The constant of integration implicit in Eq.~(\ref{eqdp}) must be
  set to the macroscopic Berry-phase $P$ of the superlattice.}
\bibitem[{exp({\natexlab{d}})}]{explan-volume}
\bibinfo{note}{In Eq.~(\ref{eqformalp}), $\Omega$ is set to 1/8 of the total
  supercell volume. More sophisticated choices would only change the $P_j$
  values by a few percent.}
\end{thebibliography}

\end{document}